\documentclass{emulateapj}
\usepackage{color}

\begin{document}
\shorttitle{Evolution of Regulus}
\shortauthors{}
\title{The Past and Future History of Regulus}

\author{S. Rappaport\altaffilmark{1}, Ph. Podsiadlowski\altaffilmark{2}, and I. Horev\altaffilmark{3}}

\altaffiltext{1}{37-602B, M.I.T. Department of Physics and Kavli
  Institute for Astrophysics and Space Research, 70 Vassar St.,
  Cambridge, MA, 02139; sar@mit.edu} \altaffiltext{2}{Department
  of Astrophysics, University of Oxford, Oxford OX1 3RH; podsi@astro.ox.ac.uk}
  \altaffiltext{3}{Visitor,
  M.I.T. Kavli Institute for Astrophysics and Space Research,
  Cambridge, MA, 02139; Physics Department, Technion -- Israel
  Institute of Technology, Haifa, Israel; inbalhorev@gmail.com} 
\begin{abstract}
We show how the recent discovery of a likely close white dwarf companion to
the well known star Regulus, one of the brightest stars in the sky, leads to 
considerable insight into the prior evolutionary history of this star, including the 
cause of its current rapid rotation.  We infer a relatively narrow range for the initial 
masses of the progenitor system: $M_{10} = 2.3 \pm 0.2~M_\odot$ and $M_{20} = 
1.7 \pm 0.2~M_\odot$, where $M_{10}$ and $M_{20}$ are the initial masses
of the progenitors of the white dwarf and Regulus, respectively.  In this scenario,
the age of the Regulus system would exceed 1 Gyr.  We also show 
that Regulus, with a current orbital period of 40 days, has an interesting future 
ahead of it.  This includes (i) a common envelope phase, and, quite possibly, (ii) 
an sdB phase, followed by (iii) an AM CVn phase with orbital periods $\lesssim 1$ hr.  
Binary evolution calculations are presented in support of this scenario. We also 
discuss alternative possibilities, emphasizing the present uncertainties in binary 
evolution theory. Thus, this one particular star system illustrates many different 
aspects of binary stellar evolution.

\end{abstract}


\section{Introduction}
\label{sec:intro}

\subsection{Properties of the Regulus System}

Regulus ($\alpha$ Leo; HD 87901) was recently discovered to be a
spectroscopic binary which has a likely white dwarf companion of mass
$M_{\rm wd}\simeq 0.3\,M_\odot$ and an orbital period of $P_{\rm
orb}\simeq 40.11$ d (\citealp{Gies08}; hereafter ``the inner binary
system''). Regulus itself has an inferred mass of $\sim$$3.4 \pm
0.2\,M_\odot$. This very bright star has been known for many years to
have at least two other companions (BC) which together form a binary system
\citep{McAl05}. This BC subsystem is located $177''$ from Regulus, too
great a distance to have ever directly interacted with the star. The B
component ($\alpha$ Leo B; HD 87884) is a $\sim$$0.8 \,M_\odot$ star of
spectral type K2\,V; the C component is a very faint M4\,V star with a
mass of $\sim$$0.2 \,M_\odot$.  The Washington Double Star Catalog
(Mason et al.~2001) lists a D component of the system, also having a 
common proper motion with the system and a separation of $217''$ from 
component A.

Located at distance of $24.3\pm0.2$ pc, Regulus is a star of spectral
type B7\,V with an apparent magnitude of 1.36, making it one of the
brightest stars in the sky. The star rotates very rapidly, with a
rotational period of $15.9$\,h, constituting $\sim$$86\%$ of its
break-up speed \citep{McAl05}. Its rapid rotation causes it to be
highly flattened, with an equatorial diameter $32\%$ greater than its
polar diameter ($4.16\,R_\odot$ and $3.14\,R_\odot$, respectively;
\citealp{McAl05}).

There is some debate regarding the age of Regulus. \citet{Ger01}
estimate it to be $\sim$$150$\,Myr old; however, a comparison of its
age with that of component B, assumed to be 
coeval, results in a $100$ Myr discrepancy. It is worth noting that
the age estimate for Regulus is based largely on the effective
temperature of the star. However, the rapid rotation causes a
difference of $\sim$$5000$ K between the poles and the equator
\citep{McAl05}, making this assessment somewhat suspect. Moreover,
this estimate assumes that Regulus, with $M \simeq 3.4\,M_\odot$, has evolved as a
single star. As we will show in this paper, the original mass of
Regulus was almost certainly much lower, and Regulus has attained its
present mass by mass transfer from its close companion. If this scenario
is correct, it is difficult to escape the conclusion that the entire Regulus
system has a probable age of $\gtrsim 1$\,Gyr. 

In this paper we use the fact that Regulus has a white dwarf companion
of close to $0.3\,M_\odot$ in a 40-day orbit to reconstruct an
approximate past history for the system (\S 2).  Based on the current
state of the Regulus inner binary, we discuss various possible future 
channels for the system (\S 3) with one of the most interesting possibilities
that it will become an AM CVn binary with a period as short as a few minutes.

\section{Prior Evolution of the Inner Regulus Binary}
\label{sec:evolution}

\subsection{$P_{\rm orb} - M_{\rm wd}$ Relation}
\label{sec:PM}

It has been known for a long time that accretion from a companion star
can result in very high rotation rates for the accretor.  However, in
the case of Regulus, no close companion star was known to exist until
the discovery by \citet{Gies08} that Regulus has a companion in a
40-day orbit that is likely a white dwarf.  \citet{Gies08} suggested
that matter accreted by Regulus from the envelope of the progenitor of
the putative white dwarf was responsible for the high rotation rate.
Since only a mass function is measured for the Regulus inner binary,
the mass of the unseen companion star is formally limited to $M
\gtrsim 0.3\,M_\odot$.  \citet{Gies08} argue against a neutron star
companion on the grounds that a $\sim$$1.4\,M_\odot$ companion would
require an orbital inclination of $i \lesssim 15^\circ$ which has only
a small a priori probability of $\sim$3\%.  No formal limit on the
orbital eccentricity is given by \citet{Gies08}, but an inspection of
their Fig.~1 suggests that $e \lesssim 0.05$.  This makes it even more
unlikely that the unseen companion is a neutron star considering that
the loss of even a small amount of matter during the supernova
explosion that gave birth to the neutron star would undoubtedly have
led to a sizable eccentricity.  We present another argument below that
the mass of the unseen companion is almost certainly near
$0.3\,M_\odot$.

Assuming that the unseen companion to Regulus is a white dwarf, we can
show why the 40-day orbit is just what is expected on theoretical
grounds.  For stars with an initial mass $\lesssim 2.5\,M_\odot$ there is
a nearly unique relation between its properties on the giant branch
and its He core mass.  Since, for Roche-lobe filling donor stars that
are lower in mass than the accretor (or even slightly higher), the
orbital period depends only on the mass and radius of the donor star,
the orbital period at the end of mass transfer from a giant star
depends only on the core mass of the donor star.  This period$-$white
dwarf mass relation has been studied in detail by Rappaport et
al.\,(1995).  They found the following approximate analytic expression:
\begin{equation}
P_{\rm orb} \simeq 1.3 \times 10^5 M_{\rm wd}^{6.25} (1+4M_{\rm wd}^4)^{3/2},
\end{equation}
where the orbital period is expressed in days and the white dwarf mass
is in units of $M_\odot$.  Rappaport et al.~(1995) estimated a
theoretical uncertainty in the white dwarf mass of $\sim \pm 18\%$
that takes into account uncertainties in the chemical composition, the
initial mass of the parent star, and the mixing length parameter.  If we
solve this expression for $M_{\rm wd}$ with $P_{\rm orb} = 40.1$ d, we
find
\begin{equation}
M_{\rm wd} \simeq 0.28 \pm 0.05 \,\,M_\odot
\end{equation}
(full uncertainty). This is entirely consistent with the measured
value of $M_{\rm wd} \gtrsim 0.3\,M_\odot$.

If we assume that the rotation axis of Regulus, which is measured to
be $\gtrsim 75^\circ$ with respect to the line of sight (McAlister et
al.~2005), coincides with the normal to the orbital plane, then the
orbital inclination angle of the Regulus inner binary should be $i
\gtrsim 75^\circ$.  That this is the case follows from the assumption
that the matter transferred from the progenitor of the white dwarf
companion was responsible for spinning up Regulus (see \S\ref{sec:spin}).
Therefore, the angular momentum vector of the orbit and that of the
rapidly rotating Regulus should coincide.  For $i \gtrsim 75^\circ$,
the mass function for Regulus (Gies et al.~2008) yields a mass for the
white dwarf of $M_{\rm wd} = 0.302 \pm 0.017$.  Again, this is highly
consistent with the theoretical value expected for a white dwarf
remnant in a 40-day circular orbit (see eq.~2; Rappaport et al.~1995).
Finally, we note that, according to the theoretical scenario for
forming the Regulus inner binary, the orbital eccentricity is expected
to be $e \lesssim 10^{-4}$ (see Rappaport et al.~1995, and references
therein) -- which can possibly be falsified in the
future.

\subsection{Constraints on the Primordial Binary}
\label{sec:constraints}

In the evolutionary scenario for the current Regulus inner binary, the
progenitor of the white dwarf is the more massive star, while the
secondary -- the progenitor of the current Regulus -- is somewhat less
massive.  The initial orbital period must be less than 40 days in
order for the He core of the primary not to exceed $\sim$0.3 $M_\odot$ (see
eq.~1).  When the primary overfills its Roche lobe, its envelope is
transferred stably, though not necessarily conservatively, to the
secondary.  At first, the orbit shrinks due to the transfer of mass
from the more, to the less, massive star.  Once the mass ratio reaches
unity, the orbit will start to expand.  The ratio of the
final-to-initial orbital period is given by
\begin{equation}
\frac{P_{\rm orb,f}}{P_{\rm orb,i}} = \frac{M_{\rm b,f}}{M_{\rm b,i}} \left(\frac{M_{\rm 1,f}}{M_{\rm 1,i}}\right)^{C_1} \left(\frac{M_{\rm 2,f}}{M_{\rm 2,i}}\right)^{C_2},
\end{equation}
where $C_1 = 3 \alpha(1-\beta)-3$ and $C_2 =
-3\alpha(1-\beta)/\beta-3$ (Podsiadlowski et al.~1992).  The parameter
$\beta$ is the fraction of the transferred matter retained by the
secondary, and $\alpha$ is the specific angular momentum in units of
the specific orbital angular momentum of the system carried away
by any matter lost from the system.  The notation is that ``1'',
``2'', and ``b'' stand for the primary, secondary, and the binary,
respectively, while ``f'' and ``i'' indicate final and initial states,
respectively.  The derivation of eq.~(3) involves the assumption that
the parameters $\alpha$ and $\beta$ are constant throughout the mass-transfer 
phase; this approximation seems justified in light of the other
theoretical uncertainties for this phase of the evolution.

\begin{figure}[t]
\centering
\includegraphics[width=0.46\textwidth]{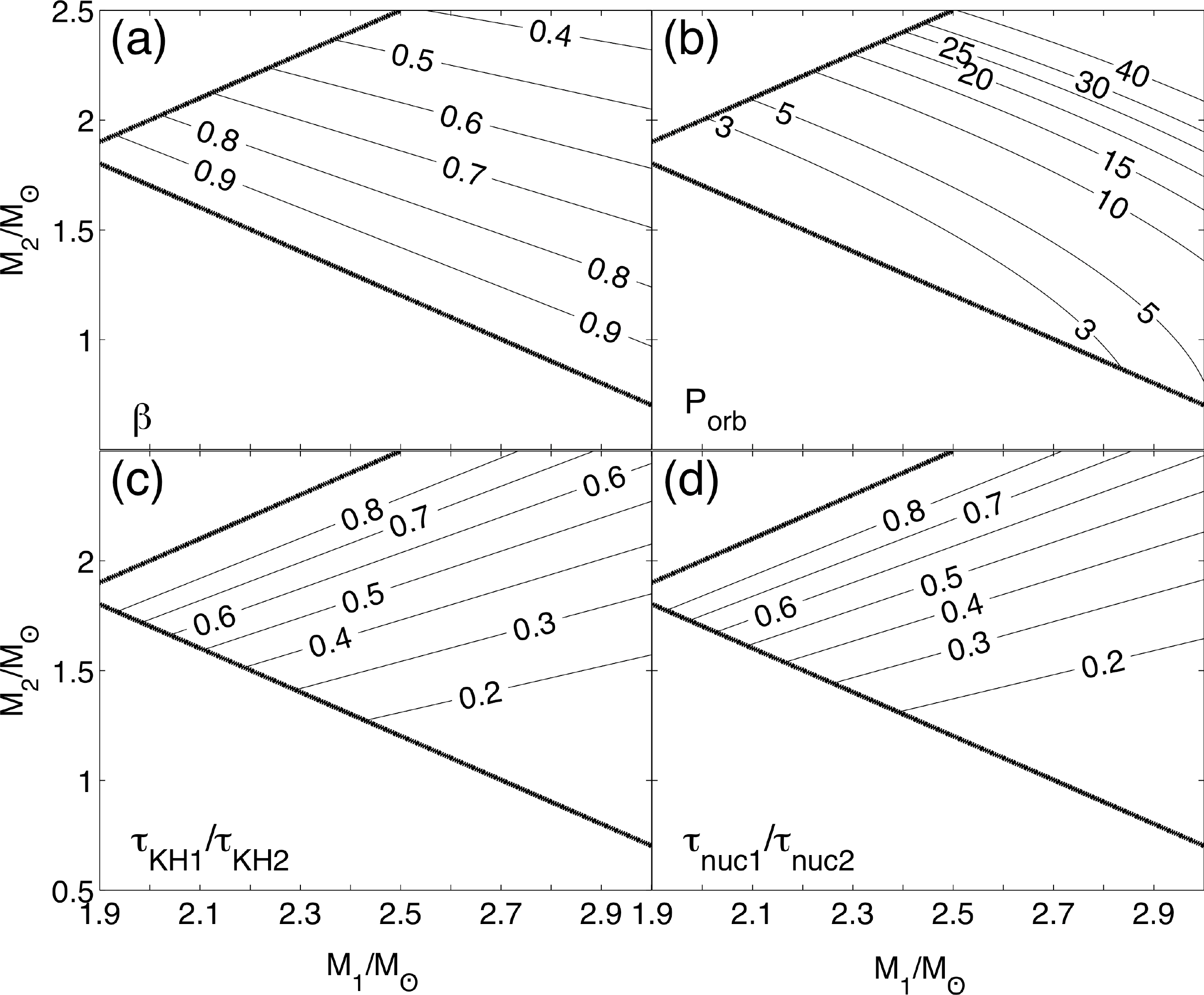}
\caption{Contours of $\beta$, $P_{\rm orb}$, $\tau_{\rm
    KH,1}/\tau_{\rm KH,2}$, and $\tau_{\rm nuc,1}/\tau_{\rm nuc,2}$ in
  the $M_{10}-M_{20}$ plane of the progenitor of the Regulus inner
  binary.  All physically realizable points in the $M_{10}-M_{20}$
  plane must lie above the heavy solid line with negative slope $-1$
  (common to all panels) in order to yield a current binary mass of
  $3.7 \pm 0.2\,M_\odot$.  Similarly, all physically meaningful points
  must lie below the heavy line of equal masses with slope $+1$.}
\label{fig:plotone}
\vspace{0.2cm}
\end{figure}

In order to study the range of possible progenitor stars for the
Regulus inner binary system, we considered primary stars of mass
$M_{10} \lesssim 3\,M_\odot$, and secondaries of any mass less than the
primary mass.  For primaries with $M_{10} \gtrsim 2.5-3\,M_\odot$ the
core mass that develops would be $M_{\rm core} \gtrsim 0.5\,M_\odot$
and would not leave a remnant consistent with the $0.3\,M_\odot$ white
dwarf that is inferred to be orbiting Regulus.  For each point in the
$M_{10}-M_{20}$ plane we estimated the ratio of thermal timescales,
$\tau_{\rm KH}$, and nuclear timescales, $\tau_{\rm nuc}$, for the
primary and secondary.  Contours of constant timescale ratios are
shown in the bottom two panels (c and d) of Fig.~1.  We take $\tau_{\rm KH}
\propto M^2/(RL)$ and $\tau_{\rm nuc} \propto M/L$, with stellar
luminosity and radius scaling like $M^{3.7}$ and $M^{0.8}$,
respectively.  We then find $\tau_{\rm KH} \propto M^{-2.5}$ and
$\tau_{\rm nuc} \propto M^{-2.7}$.

In the top left panel of Fig.~1 (panel a), we show contours of constant $\beta$,
the fraction of transferred mass retained by the secondary, in the
$M_{10}-M_{20}$ plane.  These values are inferred by comparing the
initial mass of the primordial binary with the current observed mass
of $M_{\rm b,f} \simeq 3.4 + 0.3~M_\odot = 3.7\,M_\odot$.  Finally, the
top right panel (b) in Fig.~1 shows contours of constant initial orbital
period that would lead to the observed current value of $P_{ \rm orb}
= 40$ days.  To compute $P_{\rm orb}$ we utilized eq.\,(3) with $\beta$
taken from Fig.~1a, and $\alpha$, the specific angular momentum
associated with mass loss, taken to be a constant throughout the mass 
loss process and to have a typical fiducial value of unity (but see \S\ref{sec:bincalc}).

If we require that the ratio of thermal timescales for the progenitor
stars, $\tau_{\rm KH,1}/\tau_{\rm KH,2}$ be not too different from
unity, so that the secondary can retain a sizable fraction of the
transferred matter, we can restrict the allowed range of progenitor
masses to lie above a particular contour in Fig.~1c.  We somewhat
arbitrarily choose $\tau_{\rm KH,1}/\tau_{\rm KH,2} \gtrsim 0.4$ in
order that the thermal timescales not be too disparate.  Similarly, we
require that the ratio of nuclear timescales for the progenitor stars,
$\tau_{\rm nuc,1}/\tau_{\rm nuc,2}$ be sufficiently different from
unity, so that the current Regulus has not already evolved up the
giant branch. Since, as we show in \S2.5, the initial secondary star
will spend $\sim$200 Myr with sufficient accreted mass to exceed the
mass of the original primary, and subsequently, another $\sim$50 Myr
with much more mass than the primary, as it approaches its current
mass of $3.4\,M_\odot$, the original mass of the secondary cannot be
too close to that of the primary.  In order for Regulus not to be more
evolved than its current state, we require that $\tau_{\rm
nuc,1}/\tau_{\rm nuc,2} \lesssim 0.6$, and we adopt this as a rough
upper boundary in the bottom right panel of Fig.~1.

\begin{figure}[t]
\centering
\includegraphics[width=0.48\textwidth]{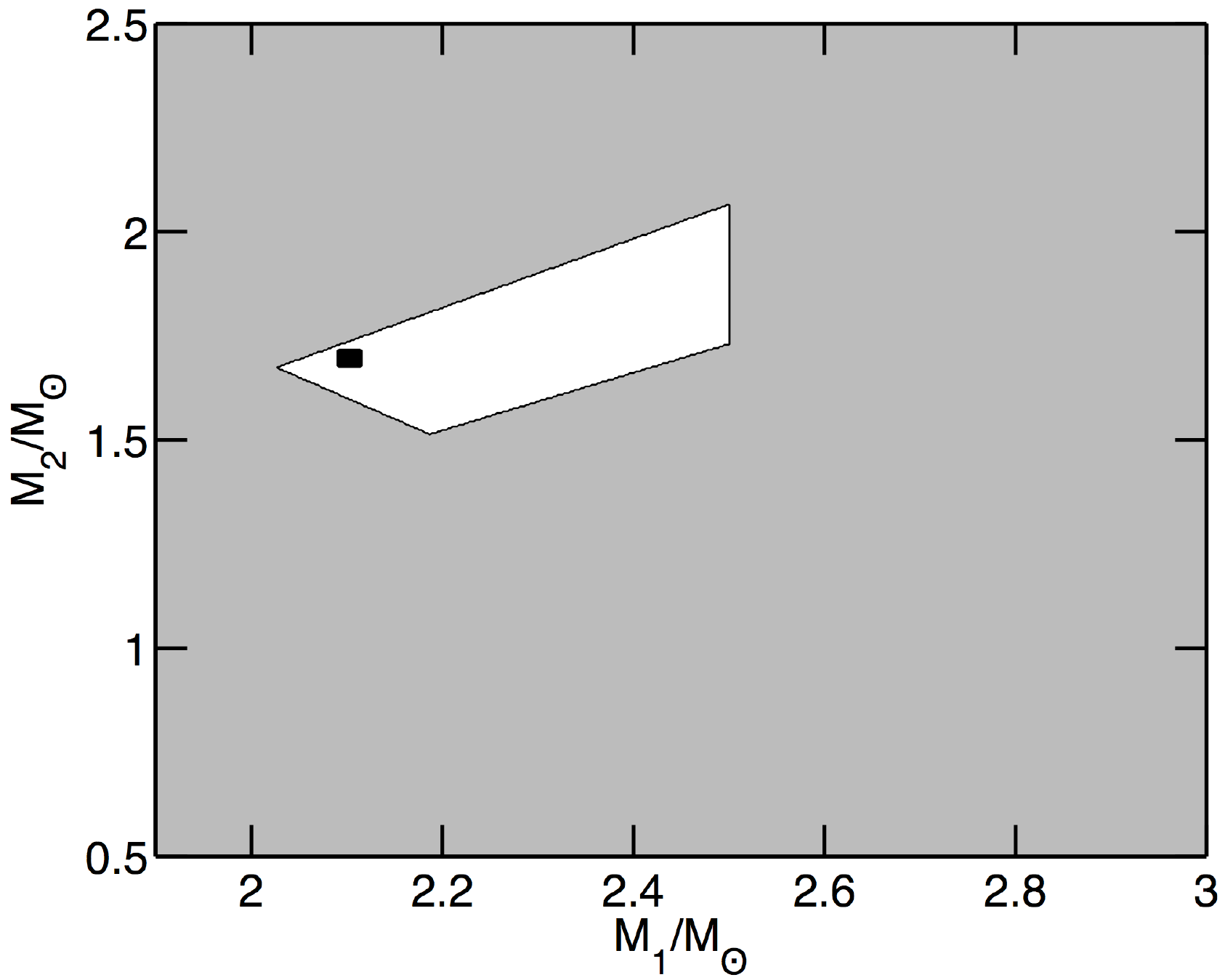}              
\caption{Allowed regions in the $M_{10}-M_{20}$ plane for the
  progenitor of the Regulus inner binary (see Fig.~1).  The cutoff for
  $M_{10} > 2.5\,M_\odot$ indicates the highest mass primary that
  could yield a white dwarf mass as low as $0.3\,M_\odot$.  The black
  square indicates the initial masses used in the numerical evolution model
  that produced the best match to the current Regulus (inner binary) system.}
\label{fig:plottwo}
\end{figure}

\subsection{Parameters of the Primordial Binary}
\label{sec:results}

Combining these constraints yields the allowed range of $M_{10}$ and
$M_{20}$ (see Fig.~2).  The most likely primordial masses are $M_{10}
\simeq 2.3 \pm 0.2\,M_\odot$ and $M_{20} \simeq 1.7 \pm 0.2\,M_\odot$.
The initial orbital period could have been anywhere in the range of
$P_{\rm orb} \simeq 1-15$ days.  Calculations of grids, and even
entire populations, of such mass transfer binaries are given, e.g., in
Nelson \& Eggleton (2001) and Willems \& Kolb (2004).  These authors
considered only conservative mass transfer, but their detailed
evolutionary calculations are nonetheless quite instructive.

\medskip

\subsection{The Origin of the Rapid Rotation}
\label{sec:spin}

It has long been argued that one way to produce a very rapidly
rotating star (e.g., a Be star) is by the accretion of mass and
angular momentum from a companion star (e.g., Pols et al.\ 1991),
similar to the origin of the Be phenomenon in Be-/X-ray binaries
(Rappaport \& van den Heuvel 1982). Generally, a star has to accrete
$\lesssim 10$\,\% of its initial mass to be spun up to near critical
rotation\footnote{We note that any of the initial masses specified in
  Figs.~1 and 2, transfer more than the requisite amount of mass to
  spin up Regulus.}  (Packet 1981). But, if the companion is a
hard-to-detect degenerate object, it is generally difficult to verify
such a mass-transfer scenario (Pols et al.\ 1991). The discovery of
the close degenerate companion to Regulus confirms the mass-transfer
hypothesis in this case. Considering that this a nearby, bright star,
it also demonstrates how difficult it is to test this scenario in
individual cases. Indeed, the majority of intermediate-mass stars 
rotating near breakup may have such an unseen companion, suggesting 
that further observational scrutiny is warranted.

\subsection{Illustrative Binary Evolution Calculations}
\label{sec:bincalc}

In order to check more quantitatively our proposed evolutionary
scenario for the formation of the current Regulus system, we carried
out a number of binary evolution calculations using a Henyey stellar
evolution code.  All calculations were carried out with an up-to-date,
standard Henyey-type stellar evolution code (Kippenhahn, Weigert, \&
Hofmeister 1967), which uses OPAL opacities (Rogers \& Iglesias 1992)
complemented with those from Alexander \& Ferguson (1994) at low
temperatures. We use solar metallicity ($Z = 0.02$), take a mixing
length of 2 pressure scale heights, and assume 0.25 pressure scale
heights of convective overshooting from the core, following the
calibration of this parameter by Schr\"oder, Pols, \& Eggleton (1997)
and Pols et al.~(1997) (for more details see Podsiadlowski, Rappaport,
\& Pfahl 2002).

We ran binary evolution sequences for a substantial number of initial
mass points lying inside the white region of Fig.~2.  In most, but
not all, cases we adopted a value for the angular momentum loss parameter
of $\alpha = 1$.  The mass retention fraction, $\beta$, is fixed from the
initial masses and the current binary mass.  Since the values of
$\beta$ typically lie between $\sim0.7-1.0$, and therefore not much
mass is lost, our results are not highly sensitive to the exact choice
for $\alpha$.  We have also verified this by running a number of
evolutionary models where $\alpha = 0.5$ and $\alpha = 1.5$.  The main
effect of adopting higher values of $\alpha$ is that the initial
orbital period can be longer, and this tends to push the onset of mass
transfer more into the regime of early case B mass transfer.

\begin{figure}[t]
\centering
\includegraphics[width=0.48\textwidth]{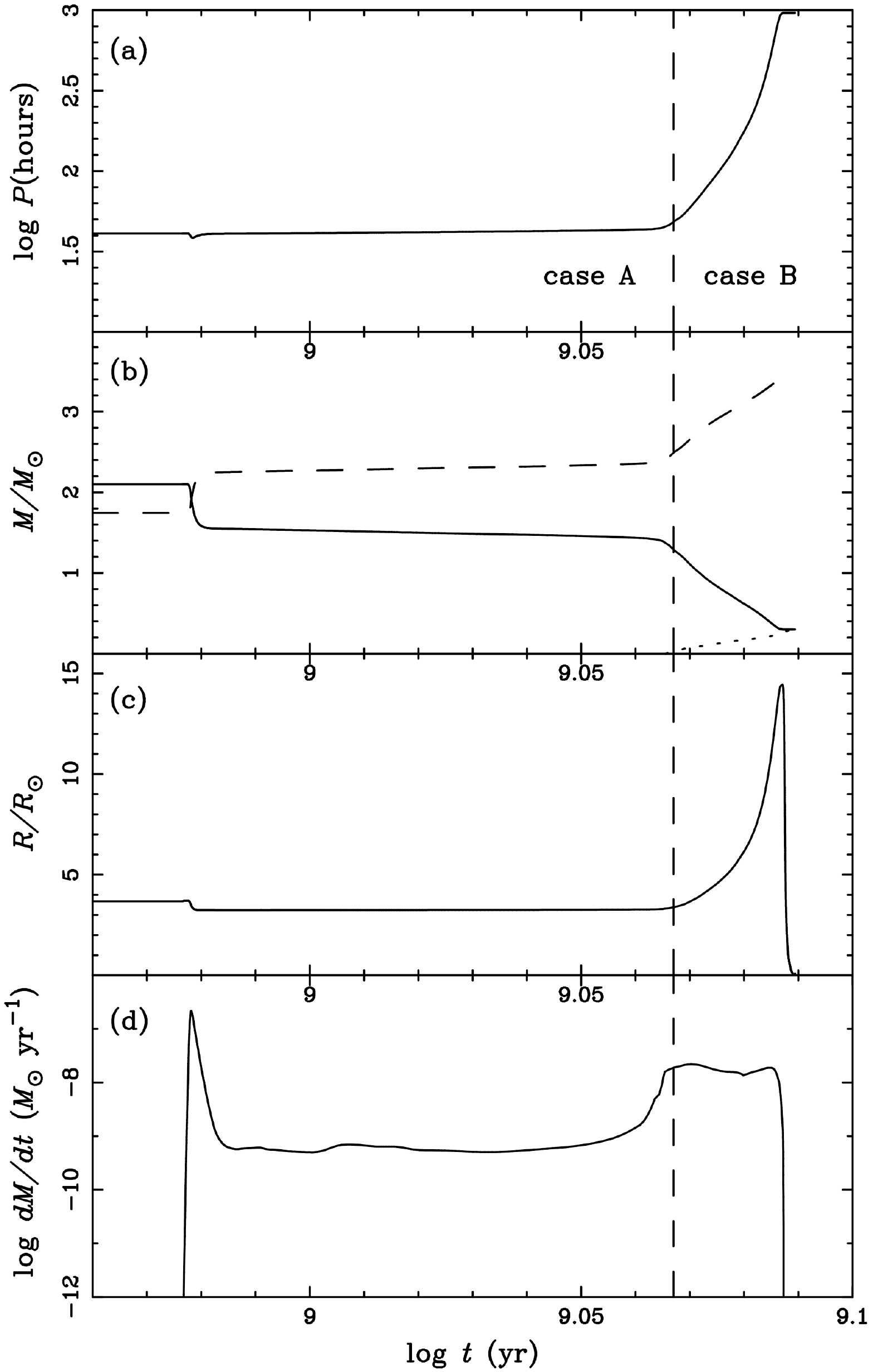}              
\caption{Evolutionary parameters as a function of time for the most
  promising model for the current Regulus system.  {\em Panel a:} the
  evolution of $P_{\rm orb}$; {\em panel b:} the masses of Regulus
  (long-dashed curve), the primary progenitor of the white dwarf (solid curve), 
  and the He core of the primary (short-dashed curve); {\em panel c:} the
  radius of the primary, the progenitor of the white dwarf; and {\em panel d:} the
  mass transfer rate from the primary.}
\label{fig:bestreg}
\vspace{0.2cm}
\end{figure}

\begin{figure}[t]
\centering
\includegraphics[width=0.47\textwidth,angle=-00]{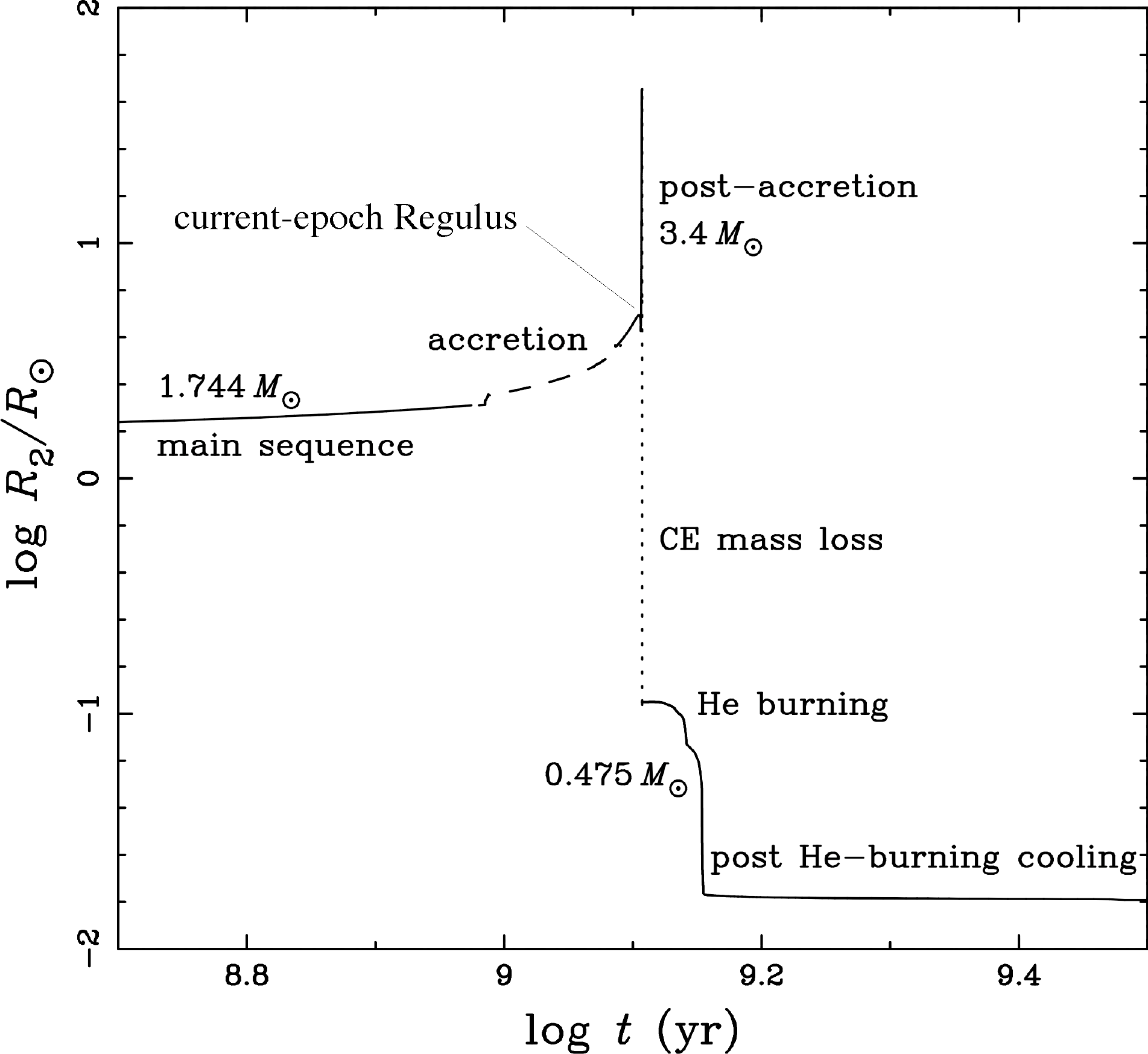}
\caption{Global view of the evolution of the radius of Regulus,
  starting from an epoch just before it starts to accrete matter from
  its companion, the accretion phase (where it reaches its current
  mass of $\sim$3.4 $M_\odot$), its subsequent evolution up the giant
  branch, followed by a common envelope (CE) phase. If the system does not
merge during the CE phase or start mass transfer shortly thereafter (see \S\ref{sec:future}),
it then becomes a 0.475\,$M_\odot$ core helium burning subdwarf and finally
a HeCO white dwarf.}
\label{fig:regev}
\vspace{0.3cm}
\end{figure}

The model which most closely matches the current-day properties of the
Regulus system has: $M_{10} = 2.1\,M_\odot$, $M_{20} = 1.74\,M_\odot$,
and $P_{\rm orb} = 40$ hours (1.7 d).  This particular choice of
initial system parameters is shown as a filled square symbol within
the white region of Fig.~\ref{fig:plottwo}.  The first of two sets of results from
the binary evolution calculation for this model is shown in Fig.~\ref{fig:bestreg}.
The four panels show (a) the evolution of $P_{\rm orb}$; (b) the progenitor
masses of Regulus (long-dashed curve) and the white dwarf progenitor 
(solid curve); (c) the radius of the primary (WD progenitor); and (d) the mass
transfer rate, $\dot M$, onto the progenitor of Regulus.  Note that as
$P_{\rm orb}$ approaches 40 days ($\sim$1000 hr), the primary has
transferred most of its envelope mass to the progenitor of Regulus,
which by then has a mass of $\sim$3.4 $M_\odot$.  At the same time,
the primary has developed a $\sim$0.3 $M_\odot$ degenerate He core 
(short-dashed curve in Fig.~3, panel b) --
the progenitor of the current white dwarf in the system\footnote{If 
the progenitor of the white dwarf had been slightly more massive
and, in particular, the He core mass had exceeded $\sim
0.32\,M_\odot$ (which would still be consistent with the present-day measured 
mass function), helium would have been ignited non-degenerately in the core 
(see \S\ref{sec:RegWD}).}. 
Fig.~\ref{fig:regev}
shows the evolution of the radius of Regulus during the epoch when it
accretes mass from the primary.  Note that the radius of Regulus, even
during the interval when it is accreting at
$\sim$$10^{-8}\,M_\odot\,{\rm yr}^{-1}$, never exceeds $4\,R_\odot$
because the mass transfer timescale is substantially longer than the
Kelvin-Helmholtz timescale.

\subsection{Status of the White Dwarf in the Regulus System}
\label{sec:RegWD}

If the $\sim$0.3 $M_\odot$ companion to Regulus, originally the core of the primordial
primary in the system, has a mass below $\sim$0.31 $M_\odot$, then the star
will not burn He to CO and will cool to become a degenerate dwarf.  After a time of several
hundred Myr, the maximum time allowed since the current Regulus binary has
been in existence, such a $\sim$0.3 $M_\odot$ white dwarf would have cooled 
to $T_{\rm eff} \lesssim$15,000 K and a luminosity of $\lesssim$0.02 $L_\odot$
(see, e.g., Althaus, Serenelli, \& Benvenuto 2001). This is consistent with the limits 
on the optical and UV emission emission from such a companion in the presence of 
Regulus, with a temperature nearly this high and a luminosity $\sim$$10^4$
times higher.

If, on the other hand, the companion to Regulus has a mass just slightly in 
excess of that required for a He star to burn He to CO in thermal equilibrium, i.e., 
$M \simeq 0.32~M_\odot$, then it would still be undergoing nuclear burning
at the current epoch.  The properties of such a low-mass He star are: $R \simeq 
0.056~R_\odot$, $L \simeq 1.4~L_\odot$, and $T_{\rm eff} \simeq 26,000$ K; and the
nuclear lifetime is in excess of 1 Gyr (see, e.g., Hurley, Pols, \& Tout 2000).  Even
though such a companion would have a substantially larger luminosity than
its degenerate counterpart, it would still not contribute very much light to the
Regulus system (i.e., $\lesssim 1\%$) for wavelengths longward of 1700 \AA.
It even seems unlikely that such a He star would have been detected in the study
of B stars (including Regulus) by Morales et al.~(2001) using the {\em IUE}
satellite and {\em EURD} spectrograph on {\em MINISAT-01}.


\subsection{Age of the Regulus System}

One consequence of our scenario is that the age of the Regulus system,
including any members bound to the inner binary, must exceed $\sim$900
Myr, the time required for the $\sim$$2.1\,M_\odot$ progenitor of the
white dwarf to evolve.  Thus, all previous age estimates of
$\sim50-200$ Myr (see, e.g., Gerbaldi et al.~2001) are far too
young.  The reason for these age discrepancies, in essence, is that 
Regulus was rejuvenated by the
accretion of nearly 1.7 $M_\odot$ of hydrogen-rich material.

\section{Future Evolution of the Inner Regulus Binary}
\label{sec:future}

\subsection{Regulus Ascending the Giant Branch}

In about 100--200 Myr, Regulus will complete its main sequence
evolution and begin its ascent of the giant branch (see
Fig.~\ref{fig:regev} for times after $10^{9.1}$ yr).  The exact time
to this phase will depend, of course, upon its current core
development.  The present orbital separation of the Regulus binary (at
$P_{\rm orb} \simeq 40$ d) is $a \simeq 74\,R_\odot$, and the
corresponding Roche-lobe radius of Regulus is $R_L \simeq 44 \pm 4.7\,R_\odot$.
By the time Regulus fills its Roche lobe, starts mass transfer to its white dwarf
companion, and the common-envelope phase commences, it will have 
developed a $\sim0.48\,M_\odot$ He core\footnote{We note that whether 
Regulus fills its Roche lobe
during or after helium core burning (when it has already developed
a CO core) depends critically on the amount of convective overshooting.
For a significantly lower value than our preferred value of
0.25 pressure scale heights, Regulus would only fill its Roche lobe
after helium core burning, by which point the hydrogen-exhausted core 
would be considerably more massive ($0.7-0.8~M_\odot$).}.
This is the core that is unveiled after
the common envelope phase (see the following section).

Note, however, that since a $3.4\,M_\odot$ star will ignite helium in
its core, its post-CE remnant will continue to burn helium. In the case of 
a successful common-envelope ejection (see \S\ref{sec:CE}) it will
appear as an sdB star in a very short-period binary for $\sim 10^8\,$yr
(e.g., Maxted et al.\ 2001; Han et al.\ 2002), and ultimately produce
a hybrid HeCO white dwarf (see \S\ref{sec:coreev}).

\subsection{Common Envelope Phase}
\label{sec:CE}

Once mass transfer from the evolved Regulus to its white dwarf
companion commences, it will quickly become dynamically unstable (due
to the extreme mass ratio of the two stars), and a common envelope
phase will ensue.  The relation between the initial orbital separation
of $a_i \simeq 76\,R_\odot$, and the post common envelope separation,
$a_f$ is given by:
\begin{equation}
\left(\frac{a_{\rm f}}{a_{\rm i}}\right) \simeq \frac{M_{\rm core,Reg}
  M_{\rm wd}}{M_{\rm Reg}} \left(M_{\rm wd}+\frac{2M_{\rm
    env,Reg}}{\lambda ~\eta~ r_{\rm L1}}\right)^{-1},
\end{equation}
where $\lambda$ is the inverse binding energy of the core of Regulus
with its envelope, in units of $R_{\rm Reg}/(GM_{\rm core,Reg}M_{\rm
env,Reg}$), $\eta$ is the fraction of the orbital binding energy that
goes into ejecting the common envelope, and $r_{\rm L1}$ is the size
of the Roche lobe of Regulus in units of the initial orbital
separation (see, e.g., Webbink 1984; Pfahl, Rappaport, \&
Podsiadlowski 2003).  For the parameters of the Regulus-white dwarf
binary, this expression yields $a_{\rm f}/a_{\rm i} \simeq 0.0046
\lambda \eta$.  The corresponding post-common envelope orbital period
of the binary, now consisting of a $\sim0.5\,M_\odot$ HeCO dwarf and a
$0.3\,M_\odot$ white dwarf, is $P_{\rm orb} \simeq 40\,(\lambda
\eta)^{3/2}$ minutes.  For plausible values of $\lambda \eta \simeq 0.1
- 1$ (see, e.g., Dewi \& Taurus 2000; 2001; Podsiadlowski,
Rappaport, \& Han 2003), the post-CE orbital periods would likely range
between $\sim$2 and 40 minutes, respectively --- if both of the compact stars were
degenerate.  Given that the HeCO star is expected to still be burning
He at the end of the common-envelope phase, and that its radius would
be no smaller than $\sim$$0.1~R_\odot$, the actual range of post-CE
orbital periods is probably limited to between $\sim$40 and $\sim$20 minutes
if a merger of the cores is to be avoided (see \S3.3.1 for more
details).  If the two dwarfs do effectively merge, they would produce
a rapidly rotating giant with unusual properties (see, \S\ref{sec:postCE}).

\begin{figure}[t]
\centering
\includegraphics[width=0.45\textwidth]{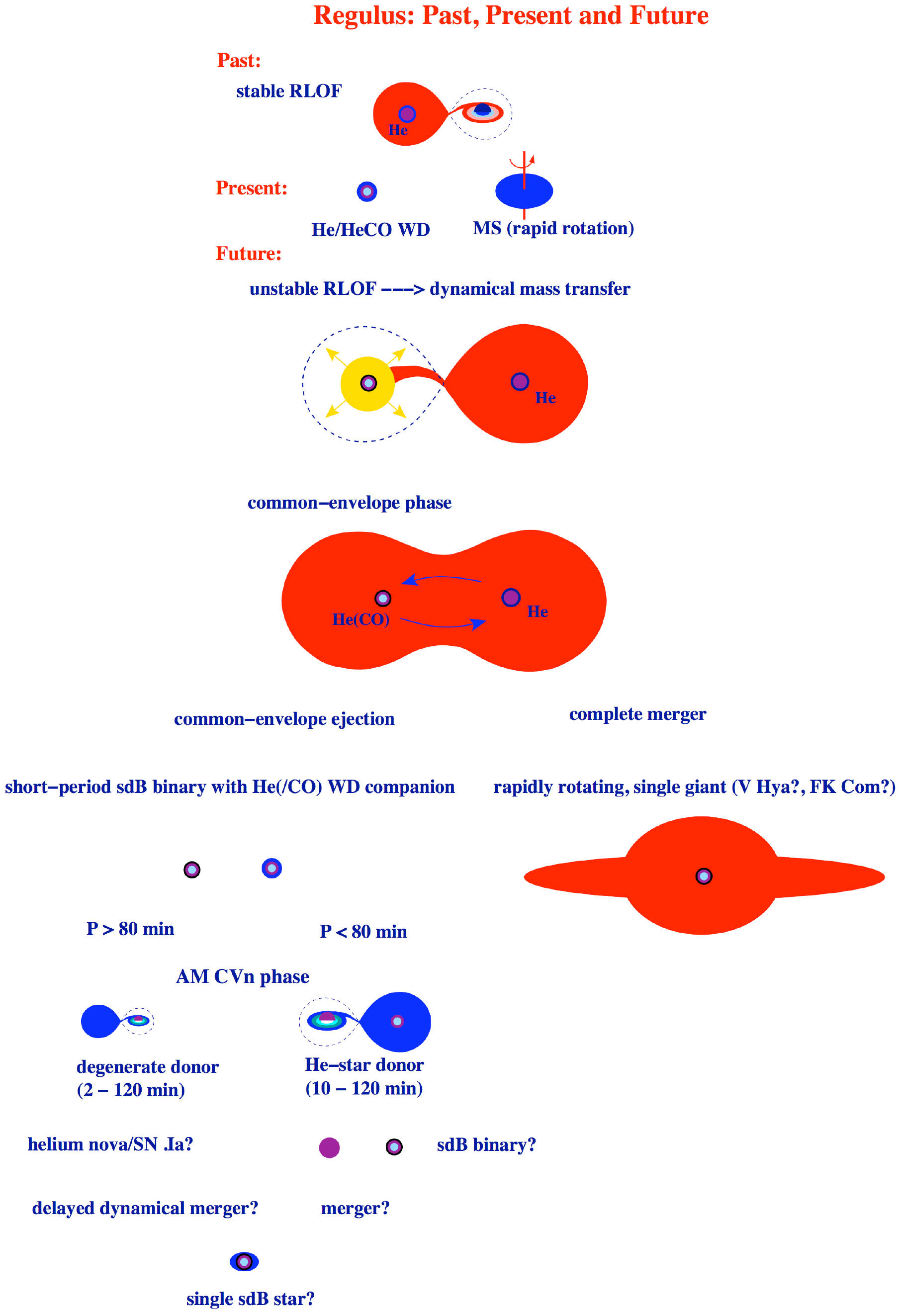}              
\caption{Evolutionary scenario: past, present, and future, for the
  Regulus inner binary.  Illustrative constituent masses and orbital
  periods corresponding to the different evolutionary phases are: (i)
  Past progenitors -- $M_{10} \simeq 2.3\,M_\odot$; $M_{20} \simeq
  1.7\,M_\odot$; $P_{\rm orb} \simeq 1-15$ days; (ii) Present --
  $M_{\rm Reg} \simeq 3.4\,M_\odot$; $M_{\rm wd} \simeq 0.3\,M_\odot$;
  $P_{\rm orb} \simeq 40$ days; (iii) Future -- unstable Roche-lobe
  overflow; starts with system parameters in (ii), but with Regulus
  having developed $\sim0.5\,M_\odot$ He core; (iv) Common Envelope
  -- spiral in of the white dwarf into the envelope of Regulus; (v)
  Post-Common Envelope phase -- leads to (a) a binary consisting of
  compact objects in an orbit with $P_{\rm orb} \gtrsim 20$ minutes (the
  former core of Regulus now appears as an sdB star), or (b) the
  system merges completely to form a rapidly rotating single giant; (vi)
  Ultracompact Binary -- depending on the post-CE envelope orbital
  period, the orbit decays due to the emission of gravitational waves
  to $P_{\rm orb}$ between $\sim$$2$ and $\sim$$20$ min, and the system 
  then appears as an AM CVn system.}
\label{fig:scen}
\vspace{0.2cm}
\end{figure}

\begin{figure}[t]
\centering
\includegraphics[width=0.48\textwidth]{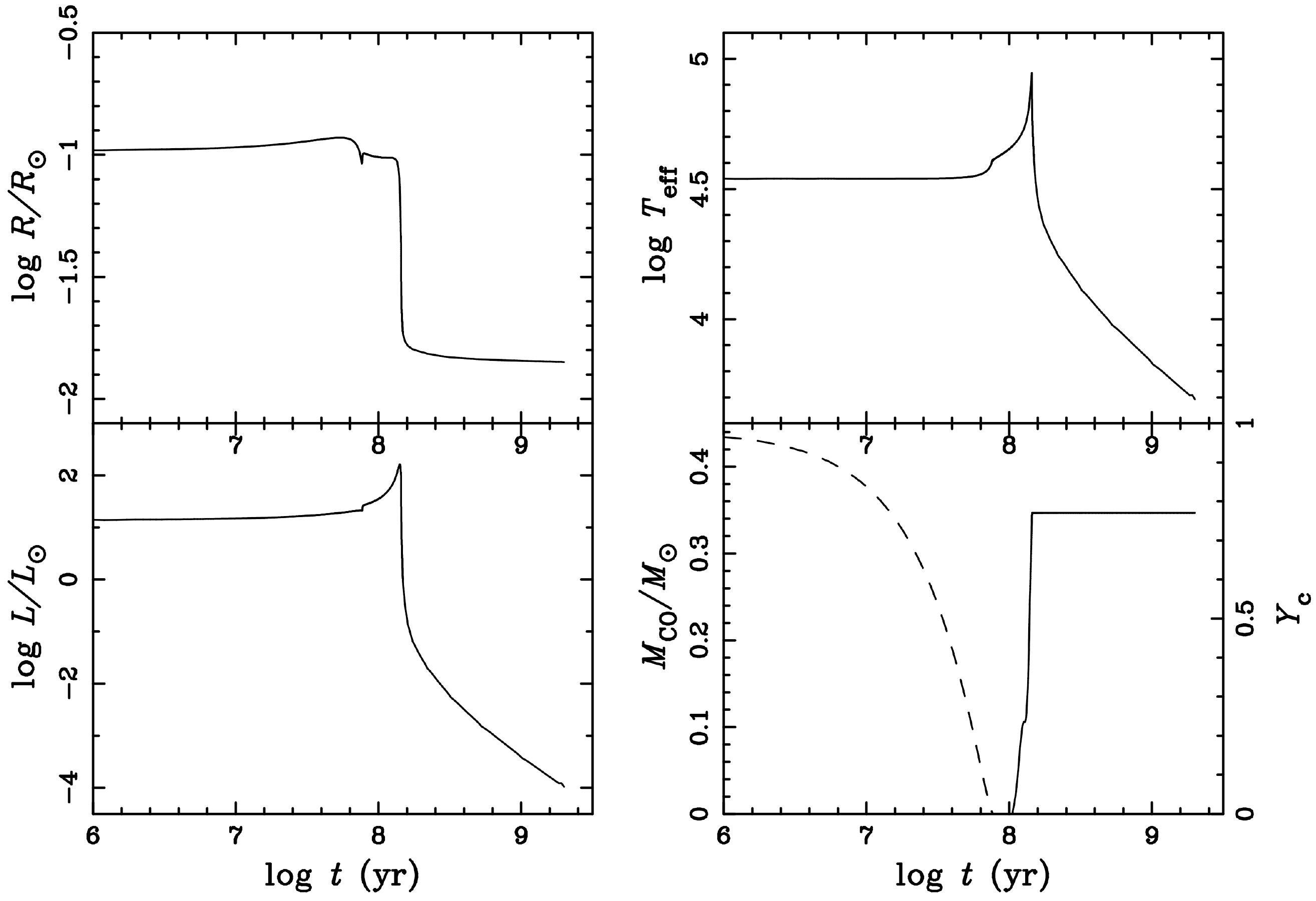}              
\caption{Evolution of an isolated He star of mass $0.475~M_\odot$.  The first three 
panels show the evolution with time of the radius, $T_{\rm eff}$, and luminosity.  
The mass of the CO core (solid curve; left scale) and the central He abundance 
(dashed curve; right scale) of the star are plotted as functions of time in
the bottom right panel.} 
\label{fig:heev}
\vspace{0.2cm}
\end{figure}

\subsection{Post Common-Envelope Phase}

\subsubsection{Possible outcomes of the CE spiral-in}
\label{sec:postCE}

After the common envelope phase, the final orbital separation should
be $a \simeq 1/3 \, \eta \lambda \, R_\odot$, where, as discussed above,
the product $\eta \lambda$ reflects the efficacy of the
common-envelope process, and is expected to be in the range of $0.1 -
1$.  What happens to the binary pair consisting of the $0.3\,M_\odot$
He white dwarf companion of Regulus and the $\sim$$
0.48\,M_\odot$ core of Regulus depends critically on the uncertain
value of $\eta \lambda$.  If the $0.3\,M_\odot$ dwarf is
degenerate, it will have a radius of $0.017\,R_\odot$, and would fit well
inside its Roche lobe, nearly independent of $\eta \lambda$.  (If it is 
still undergoing He burning, its radius would be $\sim$3 times larger.) The
$0.48\,M_\odot$ core of Regulus, on the other hand, is still burning
He to CO in its core, and will have a radius of $\sim 0.1\,R_\odot$
until the He burning phase has ended, and the HeCO remnant becomes
degenerate (see Figs.~\ref{fig:regev} and \ref{fig:heev}).  This phase lasts for 
$\sim$$100\,$Myr.

This implies that there is a minimum period of $\sim$20 min for the
post-CE system. If the ejection efficiency ($\eta\lambda$) is too low,
the orbital energy released in the spiral-in will not be able to eject
the common envelope, and the two compact objects, the 0.48\,$M_\odot$
core and the 0.3\,$M_\odot$ white dwarf, will merge inside the common
envelope.  This would likely produce a rapidly rotating single giant with possibly 
some very unusual chemical properties (perhaps similar to the unusual, 
rapidly rotating carbon star V Hydrae [Kahane et al. 1996]; also see FK Comae
stars [Bopp \& Stencel 1981] for related earlier-type counterparts).

If the common envelope is ejected, the subsequent evolution depends on
the post-CE orbital period, since it determines the timescale on which
either of the two compact components will achieve contact and start to
transfer mass again.  Immediately after a successful common-envelope 
ejection, both components (by definition) will underfill their Roche lobes, 
but the binary orbit will continue
to decay because of gravitational wave emission.   
The time for orbital decay is 
\begin{equation}
t = 68 \left[P_{\rm orb,CE}^{8/3} - P_{\rm orb}^{8/3}\right]~{\rm Myr}~~,
\end{equation}
where $P_{\rm orb,CE}$ is the orbital period (in hours) immediately after the 
common envelope, and $P_{\rm orb}$ is the orbital period at time $t$
later. The system remains a detached binary until
either of the two components starts to fill its Roche lobe.

If the immediate post-CE orbital period is $\lesssim 80\,$min,
the 0.48\,$M_\odot$ helium-burning component will start to fill its
Roche lobe while it is still burning helium in the center, and mass
transfer will start from the more massive helium-burning star to
the 0.3\,$M_\odot$ white dwarf. The system will now become an AM CVn
star with a helium-star donor (e.g., Nelemans et al.\ 2001). The stability
of the mass transfer is addressed in \S\ref{sec:amcv}.  

On the other hand, if the post-CE orbital period is longer than
$\sim$$80\,$min\footnote{Note that such periods are somewhat longer than 
anticipated from our simplistic expression for the ejection of a common envelope 
based on energetic arguments (i.e., eq.~[4]).}, the helium-burning star will have completed
helium-core-burning and become a degenerate, much smaller HeCO white
dwarf (see Figs.~\ref{fig:regev} and \ref{fig:heev}) before the system becomes 
semi-detached again. In this case, the lighter and physically larger white dwarf 
will fill its Roche lobe first, and mass transfer will take place from the lighter
white dwarf to the more massive HeCO white dwarf. The system will again appear
as an AM CVn binary (see, e.g., Nelemans et al.\ 2001), but in this case with either
a degenerate donor star with $P_{\rm orb} \simeq 2$ min and $M \simeq
0.3~M_\odot$, or with a He-burning star with $M \simeq 0.32~M_\odot$
and $P_{\rm orb} \simeq 13$ min.  

Figure~5 illustrates these different evolutionary paths in a
``scenario'' diagram. 

\subsubsection{Evolution of the exposed core of Regulus}
\label{sec:coreev}

Once the core of Regulus has been exposed after the common-envelope phase,
it will evolve as an isolated He star of mass $0.475~M_\odot$ until it or its
companion white dwarf fill its respective Roche lobe.  As mentioned above,
which star first fills its Roche lobe depends on the orbital period immediately
following the common-envelope phase.  To help understand this, we show
in Fig.\,\ref{fig:heev} the evolution of an isolated He star of mass $0.475~M_\odot$.  
The first three panels 
show the evolution with time of the radius, $T_{\rm eff}$, and luminosity.  After 
$\sim$140 Myr, the star has completed its He burning, and contracts from 
$\sim$$0.1~R_\odot$ to its final degenerate radius of $\sim$$0.015~R_\odot$.
The mass of the CO core and the central He abundance of the star are plotted 
as functions of time in the bottom right panel.  The CO core reaches a final mass
of $\sim$$0.34~M_\odot$.
	
Thus, the critical timescale for the exposed core of Regulus to complete its
nuclear burning and contract to a degenerate state is $\sim$140 Myr.  From
eq.\,(5) above we see that the original (i.e., post-CE) orbital period would 
have to be in excess of $\sim$80 minutes in order for the HeCO star to avoid
filling its Roche lobe at $P_{\rm orb} \simeq 20$ min, and for the system to shrink
via gravitational wave loses to $P_{\rm orb} \simeq 2$ min when the He white
dwarf would first overflow its Roche lobe (see \S\S\ref{sec:postCE} and \ref{sec:amcv}).
If the lower-mass He star is still undergoing He burning (e.g., for a mass of $\sim$0.32
$M_\odot$) then $P_{\rm orb}$ would be $\sim$13 min. rather than 2 min.

\begin{figure}[h]
\centering \includegraphics[width=0.47\textwidth]{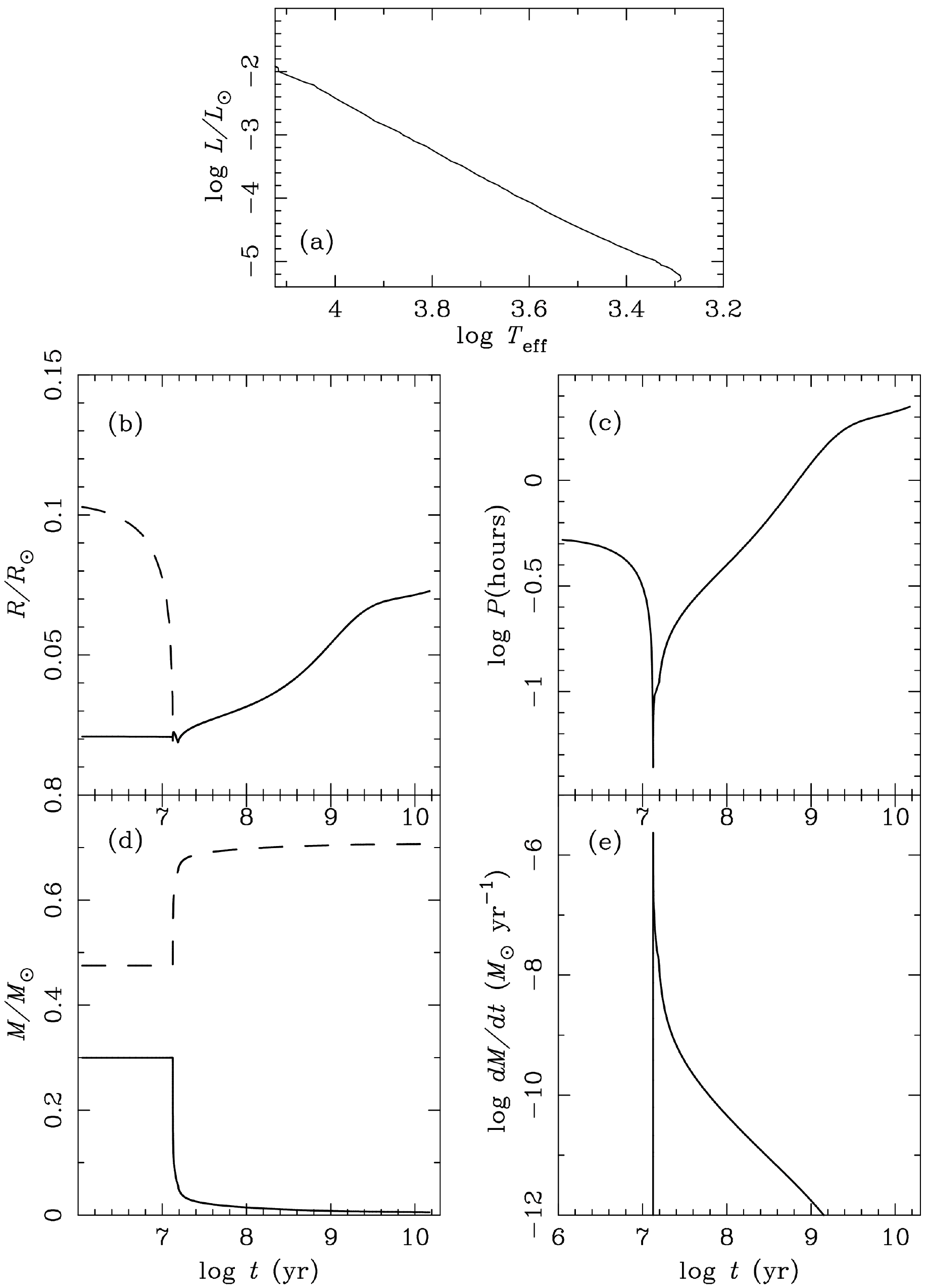}
\caption{Illustrative post-CE evolution of the $0.3\,M_\odot$ white
  dwarf and the HeCO remnant core of Regulus.  The post-CE orbital
  period was taken to be 90 minutes.  After mass transfer from the
  $0.3\,M_\odot$ white dwarf to the HeCO dwarf commences, the system
  would resemble an AM CVn system with a degenerate donor star.  
  The top panel (a) is an HR diagram for the accreting HeCO white dwarf. 
  The four panels showing the temporal evolution of the system are for
  (b) the radius of the donor star (He white dwarf, solid curve) and the Roche-lobe radius
  (dashed curve); (c) orbital period; (d) constituent masses (white dwarf of 
  initial mass $0.3\,M_\odot$, solid curve; HeCO star of initial mass $0.475\,M_\odot$, dashed curve); 
  and (e) mass transfer rate, $\dot M$.
  Note that a post-CE orbital period as long as 90 minutes is not very
  probable for the Regulus system.}
\label{fig:amcvn1}
\vspace{0.2cm}
\end{figure}

\begin{figure}[h]
\centering
\includegraphics[width=0.47\textwidth]{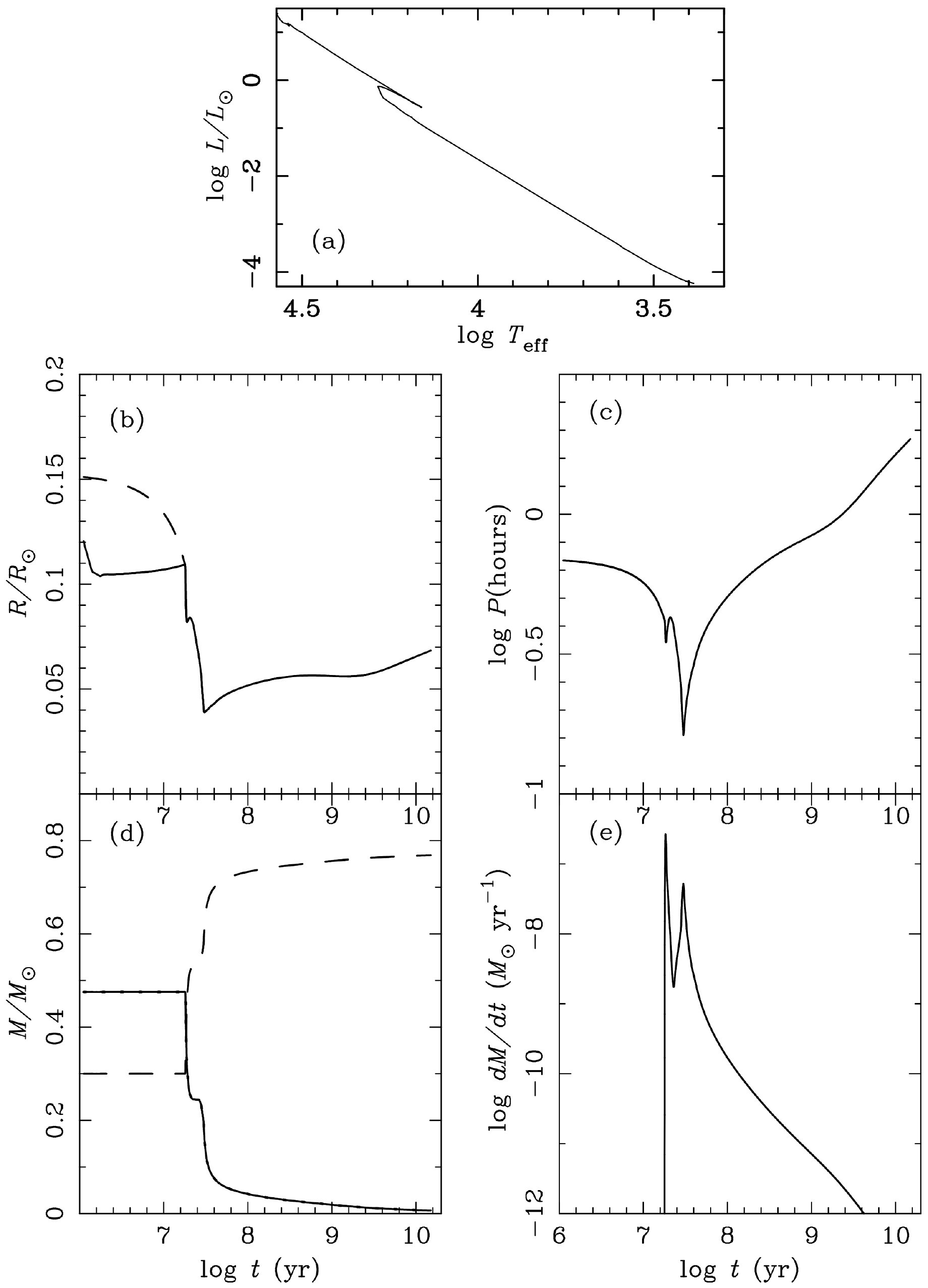}              
\caption{Illustrative post-CE evolution of the $0.3\,M_\odot$ white
  dwarf and the HeCO remnant core of Regulus.  The post-CE orbital
  period was taken to be 40 minutes.  After mass transfer from the
  $0.48\,M_\odot$ HeCO dwarf to the $0.3\,M_\odot$ white dwarf
  commences, the system would resemble an AM CVn system with a helium-star
donor.  Note that $P_{\rm orb}$ at first contact is $\sim$21 min, and then
decreases to $P_{\rm orb} \simeq 10$ min before the orbit starts to expand.
The various panels, and their description are othewise the same as in 
Fig.\,\ref{fig:amcvn1}.}
\label{fig:amcvn2}
\end{figure}

\subsubsection{Illustrative calculations of an AM CVn phase}
\label{sec:amcv}

In Fig.~\ref{fig:amcvn1} we show the evolution that results when the
post-CE orbital period is 90 min.  This allows sufficient time for the
HeCO dwarf to complete its He burning before gravitational radiation
losses bring the system into Roche-lobe contact.  The first star to
fill its Roche lobe and thereby become the donor star is the
$0.3\,M_\odot$ He white dwarf.  The initial orbital period at the start
of mass transfer is $\sim$$2\,$min (13 min.~if the low-mass He star
is a bit more massive and is still undergoing He burning). The mass 
transfer rate from the
lower to the higher mass white dwarf is stable.  It starts at a rate
of $\dot M \simeq 3 \times 10^{-6}\,M_\odot\,{\rm yr}^{-1}$ and steadily
declines over the ensuing billion years to $\lesssim
10^{-12}\,M_\odot\,{\rm yr}^{-1}$.  The details of the evolution of the
constituent masses, $P_{\rm orb}$, and $\dot M$ can be seen in
Fig.~\ref{fig:amcvn1}.  The later parts of the binary evolution can,
in fact, be computed semi-analytically (see, e.g., Rappaport, Verbunt,
\& Joss 1983; Eggleton 2007).  We find $P_{\rm orb} \simeq 40
\,t_8^{3/11}$ minutes, and $\dot M \simeq 1.8 \times 10^{-10}
t_8^{-14/11}\,M_\odot$ yr$^{-1}$, where $t_8$ is the time in units of
$10^8$ years from the start of mass transfer.  Such a system would be 
observed as an AM CVn star (see, e.g., Nelemans 2001).

In Fig.~\ref{fig:amcvn2} we show the evolution that results when the
post-CE orbital period is 40 min. In this case, the orbital decay time
is sufficiently short that the post-CE $\sim$$0.48~M_\odot$-helium star 
does not have
sufficient time to complete He burning and cool to a degenerate state
where its radius is smaller than that of the $0.3\,M_\odot$ He white
dwarf. In this case, mass transfer starts when the helium star is
still burning helium in its core (i.e., in its hot subdwarf phase),
and mass is transferred from the hot subdwarf to the $0.3\,M_\odot$ He
white dwarf. Initially, mass transfer proceeds on the thermal
timescale of the helium star due to the fact that the donor is the
more massive star.  However, after an initial phase of very high $\dot
M$, the mass ratio quickly inverts, and the subsequent mass transfer
is driven entirely by the angular momentum loss due to gravitational
radiation. Note that, nuclear burning turns off soon after the helium
star has lost a significant amount of mass (by this stage, the central
helium abundance has been reduced from 0.98 to 0.74, causing the
shrinking of the star and the associated dip in the mass-transfer
rate). The subsequent evolution of the binary is similar to that in
Fig.~\ref{fig:amcvn1}, except that the now degenerate donor star is
helium-depleted and oxygen- and carbon-enriched.

\subsubsection{The final fate of the system}
\label{sec:nuc}

For the binary evolution calculations shown in Figs.~\ref{fig:amcvn1}
and \ref{fig:amcvn2} any nuclear burning on the accreting star was
suppressed. Unfortunately, the details of the accretion and the amount
of mass lost from the system in the process of accretion and nuclear 
burning are highly uncertain, and at
this stage we can only speculate on the final evolution. In the AM CVn
case with a degenerate donor (Fig.~\ref{fig:amcvn1}), the helium in the accreting
HeCO white dwarf will at some point reignite in the helium shell under
degenerate conditions, producing a helium nova or even a mild
thermonuclear runaway (possibly producing a ``.Ia supernova'';
see Bildsten et al.\ 2007). If helium burning continues afterwards,
the accretor may swell up significantly beyond its Roche lobe, and the
system may merge completely, producing a single helium-burning hot
subdwarf (most likely a helium-rich sdO star; Stroer et al.~2007).

In the case of an AM CVn binary with a non-degenerate helium donor
(Fig.\,\ref{fig:amcvn2}), the accreting
white dwarf will experience a central helium flash when its mass reaches
the critical helium-flash mass ($\sim$$0.48\,M_\odot$). The system
will probably survive the helium flash, although the system
may become detached, and the accretor will now become a hot subdwarf.

\section{Discussion}
\label{sec:discuss}

We have discussed quantitatively the interesting past and future
history of Regulus.  The entire evolutionary scenario is summarized in
Fig.~\ref{fig:scen}.  We have shown how the present 40-day
orbital period of Regulus and its $0.3\,M_\odot$ white dwarf companion
match very well the $P_{\rm orb}-M_{\rm wd}$ relation predicted by
stellar evolution theory.  The masses of the progenitor stars are
inferred to fall in the range of $M_{10} \simeq 2.3 \pm
0.2\,M_\odot$ and $M_{20} \simeq 1.7 \pm 0.2\,M_\odot$.  Our best
numerically computed model is: $M_{10} = 2.1\,M_\odot$, $M_{20} =
1.74\,M_\odot$, $P_{\rm orb} =40$ hours (1.7 d).

In the future, Regulus will undergo a common envelope phase wherein
the white dwarf will spiral into, and eject, the envelope of Regulus.
If the envelope of Regulus is too tightly bound, then the system may
merge, forming a rapidly rotating single giant with unusual properties,
possibly related to V Hya and FK Com stars.  If, on the other hand, the
common envelope halts at orbital periods between $P_{\rm orb} \simeq
20$ and 80 min, the result will be a compact binary consisting of a
HeCO dwarf of mass $0.48\,M_\odot$ which will start transferring mass
to the $0.3\,M_\odot$ white dwarf companion (i.e., the current
companion of Regulus).  For longer post-CE orbital periods, the
gravitational wave decay timescale is sufficiently long to allow the
HeCO star to complete He burning and to cool to a largely degenerate
state.  The system would then come into Roche-lobe contact at very
short orbital periods ($P_{\rm orb} \simeq 2-13\,$min).  In both of these
latter two cases, the ensuing mass transfer will likely be stable (at
least initially) and lead to an AM CVn system whose orbital period,
after reaching a minimum,
will slowly grow to $\sim$20 minutes within $\sim$20 Myr, and to $\sim$40
minutes by $\sim$200 Myr.  Nuclear burning on the accreting dwarf is
difficult to compute, and we leave this for a future exercise.
Such burning activity could cause the donor to swell up, and this, in
turn, could cause the compact binary to merge.

Currently we are seeing Regulus as an apparently ordinary star in the
middle of an extraordinary evolutionary journey.

\acknowledgements

The authors thank the referee for comments and suggestions that led to a 
greatly improved paper.  We are grateful to Norbert Langer for stimulating 
discussions.  SR received some support from Chandra Grant TM5-6003X.  
IH thanks MKI for its hospitality during her visit.  We are grateful to Josiah 
Schwab for assistance with some of the calculations.

\end{document}